\documentclass[preprint,showpacs,preprintnumbers,amsmath,amssymb]{revtex4}

\usepackage{graphicx}% Include figure files
\usepackage{dcolumn}% Align table columns on decimal point
\usepackage{bm}% bold math
\usepackage[]{ams,amsmath,amssymb,epsfig,color}
\newcommand{\be}{\begin{equation}}
\newcommand{\ee}{\end{equation}}
\begin{document}

%\preprint{JPhysB}

\title{Chaotic dynamics of two 1/2 spin-qubit system in the optical cavity}
%Manuscript Title:\\with Forced Linebreak}% Force line breaks with \\
%\title{Dynamics of dipolar molecular chain in crossed static: Soliton formation}
%Manuscript Title:\\with Forced Linebreak}% Force line breaks with \\

%
\author{ L. Chotorlishvili *,Z. Toklikishvili }

  \email{lchotor33@yahoo.com}
\affiliation{Physics Department of the Tbilisi State
University,Chavchavadze av.3,0128, Tbilisi, Georgia }
\date{\today}% It is always \today, today,

\begin{abstract}
Spin systems are one of the most promising candidates for quantum
computation. At the same time control of a system's quantum state
during time evolution is one of the actual problems. It is usually
considered that to hold well-known resonance condition in magnetic
resonance is sufficient to control spin system. But because of
nonlinearity of the system, obstructions of control of system's
quantum state may emerge.

In particular quantum dynamics of two 1/2 spin-qubit system in the
optical cavity is studied in this work. The problem under study is
a generalization of paradigmatic model for Cavity Quantum
Electrodynamics of James-Cummings model in case of interacting
spins. In this work it is shown that dynamics is chaotic when
taking into account center-of-mass motion of the qubit and recoil
effect. And besides even in case of zero detuning chaotic dynamics
emerges in the system. It is also shown in this work that because
of the chaotic dynamics the system execute irreversible transition
from pure quantum-mechanical state to mixed one. Irreversibility
in its turn is an obstacle for controlling state of
quantum-mechanical system.

\end{abstract}

%Valid PACS numbers may be entered using the \verb+\pacs{#1}+
%command.
\maketitle

%\section{\label{Sec:Introduction} Introduction}

\section{Introduction}
Cavity quantum electrodynamics (CQED) is a rapidly developing
field of physics studying the interaction of atoms with photons in
the high-finesse cavities \cite{Aoki,Mabuchi,Hood,Raimond}.
Interest to such a systems basically is caused by two facts: One
of them is the possibility of more deep understanding of quantum
dynamics of open systems. Second argument is a possibility of
practical application in the field of quantum computing
\cite{Tuchette}. In particular CQED experiments implement a
situation so simple that their results are of great importance for
better understanding of fundamental postulates of quantum theory
\cite{Wineland}. They are thus appropriate for tests of basic
quantum properties: quantum superposition \cite{Schleich},
complementarily or entanglement
\cite{Fujisaki,Scott,Novaes,Xie,Angelo}. In the context of quantum
information processing, the atom and cavity are long-lived qubits,
and there mutual interaction provides a controllable entanglement
mechanism an essential requirement for quantum computing
\cite{Mabuchi,Hood,Raimond}. In general dissipation processes must
be taken into account when discussing problems of CQED. In
particular there are two dissipative channels for systems the atom
may spontaneously emit onto modes other then preferred cavity
mode, and photons may pass through the cavity output coupling
mirror. But modern experiments in CQED have achieved strong
atom-field coupling for the strength of the coupling exceeds both
decay processes \cite{Ye,van Enk,Punstermann}. If so, then problem
is reduced to the Jaynes-Cummings (JC) Hamiltonian, which models
the interaction of a single mode of an optical cavity having
resonant frequency with a two level atom comprised of a ground and
exited states \cite{Schleich}.

One of the most promising candidates for quantum computation is
spin systems \cite{Loss,Kane,Skinner,Ladd,de Sousa}. In
\cite{Xiao-Zhong Yuan} was considered a two-spin-qubit system
interacting with bath spins via Heisenberg XY interaction. The
authors of indicated work could show that the problem is reduced
to study JC two spin model. It has turned out that dynamics is
non-Markovian. But in most general case atom- radiation field
interaction should involve not only the internal atomic
transitions and field states but also the center-of-mass motion of
the atom and recoil effect. The study of such a case is the aim of
this work. The subject of our interest is the following: it is
well known that for quantum computing exact control of the spins
system is necessary. That is why  zero detuning is a matter of
interest. In \cite{Prants} has been shown that even taking into
account of recoil effect and center-of-mass motion for zero
detuning, dynamics is regular and chaos emerges only, when
detuning is non-zero. But what will happen in case of modified two
spin JS model, it is not clear for the present. This work is
devoted to the study of this problem. The first part of this work
is devoted to quasi-classical consideration. In the second part we
shall try to give kinetic consideration of the phenomena.

\section{Quantum Nonlinear Resonance}
As was noted in the introduction we would like to consider more
general model proposed in \cite{Prants}. It is not difficult to
note that the Hamiltonian of the system of our interest
\cite{Xiao-Zhong Yuan} takes the form when taking into account
center-of-mass motion and recoil effect \cite{Prants}:
 \be
\hat{H}=\frac{\hat{p}^{2}}{2m}+\hat{H}_{S}+\hat{H}_{SB}+\hat{H}_{B},\ee
where $\frac{\hat{p}^{2}}{2m}$ is a kinetic energy of two spin
qubit system placed in the resonator. It is supposed that qubit is
composed of two spin 1/2 atoms \cite{Xiao-Zhong Yuan}. The spin
part of the Hamiltonian has the form:
 \be
\hat{H}_{S}=\omega_{0}(\hat{S}_{1}^{z}+\hat{S}_{2}^{z})+\Omega
(\hat{S}_{1}^{+}\hat{S}_{2}^{-}+\hat{S}_{1}^{-}\hat{S}_{2}^{+}),\ee
where  $\hbar=1$, $\omega_{0}$ is Zeeman frequency of the spins
being in the field inside the resonator, $\Omega$ is a constant of
dipole interaction between the spins in frequency units. The third
term in (1) presents itself spin 1/2 atoms interaction with
resonator field: \be \hat{H}_{SB}=-g_{0}\cos
(k_{f}\hat{x})((\hat{S}_{1}^{+}+\hat{S}_{2}^{+})\hat{b}+(\hat{S}_{1}^{-}+\hat{S}_{2}^{-})\hat{b}^{+}),\ee
here $g_{0}$ is amplitude value of the qubit-field coupling that
depends on the position of qubit $\hat{x}$ inside a cavity. The
last term in (1) is the Hamiltonian of the field:
 \be \hat{H}_{B}= \omega_{f}\hat{b}^{+}b, \ee
where $\omega_{f}$  is the selected frequency of radiation field,
$k_{f}$ is the wave number.

Taking into account commutation relation between operators
\cite{Landau}:$$[\hat{b},\hat{b}^{+}]=1,~~[\hat{S}_{z},\hat{S}^{\pm}]=\pm
\hat{S}^{\pm},~~[\hat{S}^{+}\hat{S}^{-}]=2\hat{S}_{z} $$ it is
possible to obtain the following Heisenberg equation of motions:
$$\frac{d\hat{x}}{dt}=\frac{\hat{p}}{m},$$
$$\frac{d\hat{p}}{dt}=-g_{0}k_{f}\sin
(k_{f}\hat{x})((\hat{S}_{1}^{-}\hat{b}^{+}+\hat{S}_{1}^{+}\hat{b})+(\hat{S}_{2}^{-}\hat{b}^{+}+\hat{S}_{2}^{+}\hat{b})),$$
$$\frac{d\hat{S}_{1}^{+}}{dt}=i\omega_{0}\hat{S}_{1}^{+}-2i\Omega\hat{S}_{1}^{z}\hat{S}_{2}^{+}+2ig\hat{S}_{1}^{z}\hat{b}^{+}\cos(k_{f}\hat{x}),$$
$$\frac{d\hat{S}_{1}^{-}}{dt}=-i\omega_{0}\hat{S}_{1}^{-}+2i\Omega\hat{S}_{1}^{z}\hat{S}_{2}^{-}-2ig\hat{S}_{1}^{z}\hat{b}\cos(k_{f}\hat{x}),$$
$$\frac{d\hat{S}_{1}^{z}}{dt}=-ig\cos(k_{f}\hat{x})(\hat{S}_{1}^{-}\hat{b}^{+}-\hat{S}_{1}^{+}\hat{b})-i\Omega(\hat{S}_{1}^{+}\hat{S}_{2}^{-}-\hat{S}_{1}^{-}\hat{S}_{2}^{+}),$$
$$\frac{d\hat{S}_{2}^{+}}{dt}=i\omega_{0}\hat{S}_{2}^{+}-2i\Omega\hat{S}_{2}^{z}\hat{S}_{1}^{+}+2ig\hat{S}_{2}^{z}\hat{b}^{+}\cos(k_{f}\hat{x}),$$
$$\frac{d\hat{S}_{2}^{-}}{dt}=-i\omega_{0}\hat{S}_{2}^{-}+2i\Omega\hat{S}_{2}^{z}\hat{S}_{1}^{-}-2ig\hat{S}_{2}^{z}\hat{b}\cos(k_{f}\hat{x}),$$
$$\frac{d\hat{S}_{2}^{z}}{dt}=-ig_{0}\cos(k_{f}\hat{x})(\hat{S}_{2}^{-}\hat{b}^{+}-\hat{S}_{2}^{+}\hat{b})-i\Omega(\hat{S}_{2}^{+}\hat{S}_{1}^{-}-\hat{S}_{2}^{-}\hat{S}_{1}^{+}),$$
$$\frac{d\hat{b}^{+}}{dt}=i\omega_{f}\hat{b}^{+}-ig_{0}\cos(k_{f}\hat{x})(\hat{S}_{1}^{+}+\hat{S}_{2}^{+}),$$
\be
\frac{d\hat{b}}{dt}=-i\omega_{f}\hat{b}+ig_{0}\cos(k_{f}\hat{x})(\hat{S}_{1}^{-}+\hat{S}_{2}^{-}).\ee
After going to the representation of interaction: \be
\hat{b}^{+}(t)=e^{i\omega_{f}t}\hat{b},~~\hat{b}(t)=e^{-i\omega_{f}t}\hat{b},~~\hat{S}^{\pm}(t)=e^{i\omega_{0}t}\hat{S}^{\pm}\ee
and introducing new variables by means of quasy-classical
averaging \cite{Prants}: \be
x=k_{f}<\hat{x}>,~~p=\frac{<\hat{p}>}{k_{f}},~~b_{x}=\frac{1}{2}<\hat{b}^{+}+\hat{b}>,~~
b_{y}=\frac{1}{2i}<\hat{b}-\hat{b}^{+}>;\ee
$$S_{1,2}^{x}=\frac{1}{2}<\hat{S}_{1,2}^{-}+\hat{S}_{1,2}^{+}>,~~S_{1,2}^{y}=\frac{1}{2i}<\hat{S}_{1,2}^{-}-\hat{S}_{1,2}^{+}>;$$
$$\alpha=\frac{k_{f}^{2}}{mg_{0}},~~\delta =\frac{\omega_{f}-\omega_{0}}{g_{0}},~~\beta=\Omega/g_{0},~~r=g_{0}t.$$
Taking into account (6), (7) we obtain from (5):
$$\frac{dx}{d\tau}=\alpha p,$$
$$\frac{dp}{d\tau}=-2\sin x((S_{1}^{x}b_{x}+S_{1}^{y}b_{y})+(S_{2}^{x}b_{x}+S_{2}^{y}b_{y})),$$
$$\frac{dS_{1}^{x}}{d\tau}=-\delta S_{1}^{y}+2S_{1}^{z}b_{y}\cos x-2\beta S_{1}^{z}S_{2}^{y},$$
$$\frac{dS_{1}^{y}}{d\tau}=\delta S_{1}^{x}-2S_{1}^{z}b_{x}\cos x+2\beta S_{1}^{z}S_{2}^{x},$$
$$\frac{dS_{1}^{z}}{d\tau}=2\cos x(S_{1}^{y}b_{x}-S_{1}^{x}b_{y})+2\beta (S_{1}^{x}S_{2}^{y}-S_{1}^{y}S_{2}^{x}),$$
$$\frac{dS_{2}^{x}}{d\tau}=-\delta S_{2}^{y}+2S_{2}^{z}b_{y}\cos x-2\beta S_{2}^{z}S_{1}^{y},$$
$$\frac{dS_{2}^{y}}{d\tau}=\delta S_{2}^{x}-2S_{2}^{x}b_{x}\cos x+2\beta S_{2}^{z}S_{1}^{x},$$
$$\frac{dS_{2}^{z}}{d\tau}=2\cos x(S_{2}^{y}b_{x}-S_{2}^{x}b_{y})+2\beta (S_{2}^{x}S_{1}^{y}-S_{2}^{y}S_{1}^{x}),$$
$$\frac{db_{x}}{d\tau}=-\cos x (S_{1}^{y}+S_{2}^{y}),$$
\be \frac{db_{y}}{d\tau}=-\cos x (S_{1}^{x}+S_{2}^{x}).\ee It is
readily seen that the equations (8) allows the following integrals
of motion: \be
S_{1,2}^{2}=(S_{1,2}^{x})^{2}+(S_{1,2}^{y})^{2}+(S_{1,2}^{z})^{2},~~N=b_{x}^{2}+b_{y}^{2}+S_{1}^{z}+S_{2}^{z},\ee
$$W=\frac{\alpha p^{2}}{2}+2\beta (S_{1}^{x}S_{2}^{x}+S_{1}^{y}S_{2}^{y})-2\cos x((S_{1}^{x}b_{x}+S_{1}^{y}b_{y})+(S_{2}^{x}b_{x}+S_{2}^{y}b_{y}))-\delta (S_{1}^{z}+S_{2}^{z}).$$
Introducing the new variables:
$$U_{1}=2(S_{1}^{x}b_{x}+S_{1}^{y}b_{y}),~~U_{2}=2(S_{2}^{x}b_{x}+S_{2}^{y}b_{y}),$$
$$\nu_{1}=2(b_{y}S_{1}^{x}-b_{x}S_{1}^{y}),~~\nu_{2}=2(b_{y}S_{2}^{x}-b_{x}S_{2}^{y}),$$
\be
g=(S_{1}^{x}S_{2}^{y}-S_{1}^{y}S_{2}^{x}),~~f=(S_{1}^{x}S_{2}^{x}+S_{1}^{y}S_{2}^{y}).\ee
Taking into account the new variables (10) and integrals of motion
(9), the set of equation (8) can be rewritten in more compact
form:
$$\frac{dx}{d\tau}=\alpha p,$$
$$\frac{dp}{d\tau}=-2\sin x(U_{1}+U_{2}),$$
$$\frac{dS_{1}^{z}}{d\tau}=-2\nu_{1}\cos x+2\beta g,$$
$$\frac{dS_{2}^{z}}{d\tau}=-2\nu_{2}\cos x-2\beta g,$$
$$\frac{dU_{1}}{d\tau}=\delta \nu_{1}+2\beta S_{1}^{z}\nu_{2}-2g\cos x,$$
$$\frac{dU_{2}}{d\tau}=\delta \nu_{2}+2\beta S_{2}^{z}\nu_{1}+2g\cos x,$$
$$\frac{d\nu_{1}}{d\tau}=-\delta U_{1}+2\cos x (S_{1}^{2}-3(S_{1}^{z})^{2}+2NS_{1}^{z}-2S_{1}^{z}S_{2}^{z}+f)-2\beta S_{1}^{z}U_{2},$$
$$\frac{d\nu_{2}}{d\tau}=-\delta U_{2}+2\cos x (S_{2}^{2}-3(S_{2}^{z})^{2}+2NS_{2}^{z}-2S_{1}^{z}S_{2}^{z}+f)-2\beta S_{2}^{z}U_{1},$$
$$\frac{dg}{d\tau}=\cos x(S_{1}^{z}U_{2}-S_{2}^{z}U_{1})-2\beta S_{1}^{z}(S_{2}^{2}-(S_{2}^{z})^{2})+2\beta S_{2}^{z}(S_{1}^{2}-(S_{1}^{z})^{2}),$$
\be \frac{df}{d\tau} = (S_{1}^{z}\nu_{2}+S_{2}^{z}\nu_{1}) \cos
x.\ee By direct checking one can be convinced, that because of
complex structure of the set (11), even for zero detuning $\delta
=0$, it is impossible to obtain analytical solution. Thus, unlike
the problem studied in \cite{Prants}, in case of qubit, taking
into account of recoil effect and center-of-mass motion leads to
nonintegrability of the problem even for zero detuning. Because of
nonlinearity of the set (11) we expect to obtain chaotic
solutions. If so, the state of qubit will not be possible to be
controlled.

We have integrated the set of equation (11) for the realistic
values of parameters from the point of view of experiment
\cite{Ye,Punstermann} $\delta
=0,~~\alpha=0.01,~~\beta=0.5,~~S_{1}^{2}=S_{2}^{2}=\frac{3}{4}$.
The results of numerical integration are presented on Fig.1,2.
\begin{figure}[t]
 \centering
  \includegraphics[width=14cm]{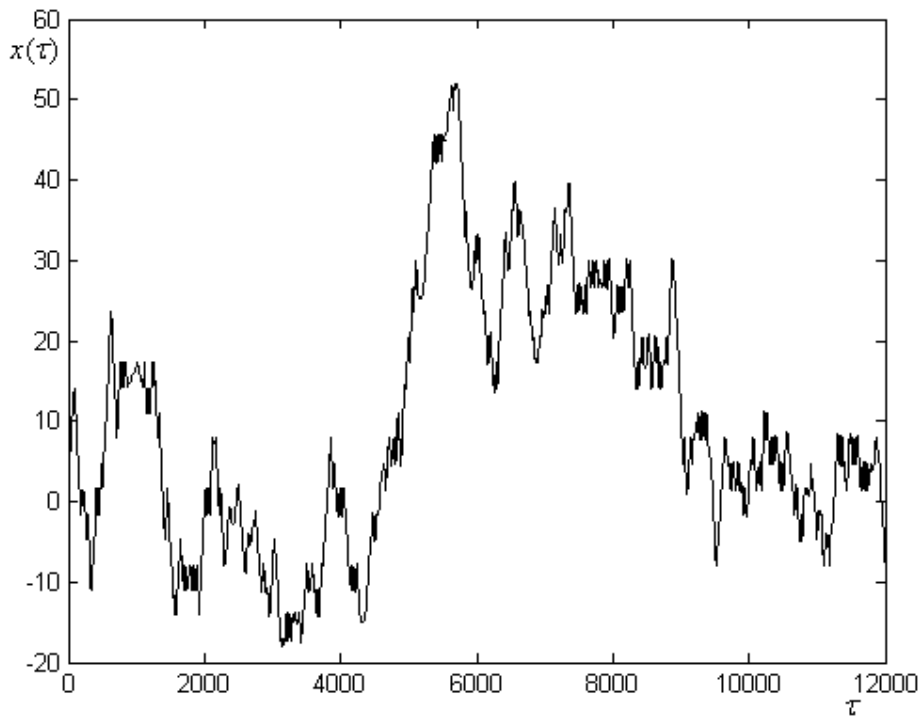}
  \caption{The graph of dependence of the system coordinates on time
$x(\tau)$. The graph is plotted for the following parameters
$x(0)=1.6,$ $p(0)=9.1,$ $S_{1}^{z}(0)=S_{2}^{z}=0.863,$
$U_{1}(0)=0.000081,$ $U_{2}(0)=0.000082,$ $\nu_{1}(0)=0.000083,$
$\nu_{2}(0)=0.000084,$  $g(0)=0.0000845,$ $f(0)=0.0000846$. As is
seen from the plot trajectory has the chaotic form.}\label{fig:1}
\end{figure}

 As is seen from Fig.1, the dynamics of the system even for zero
detuning $\delta =0$ has chaotic form.

\begin{figure}[t]
 \centering
  \includegraphics[width=14cm]{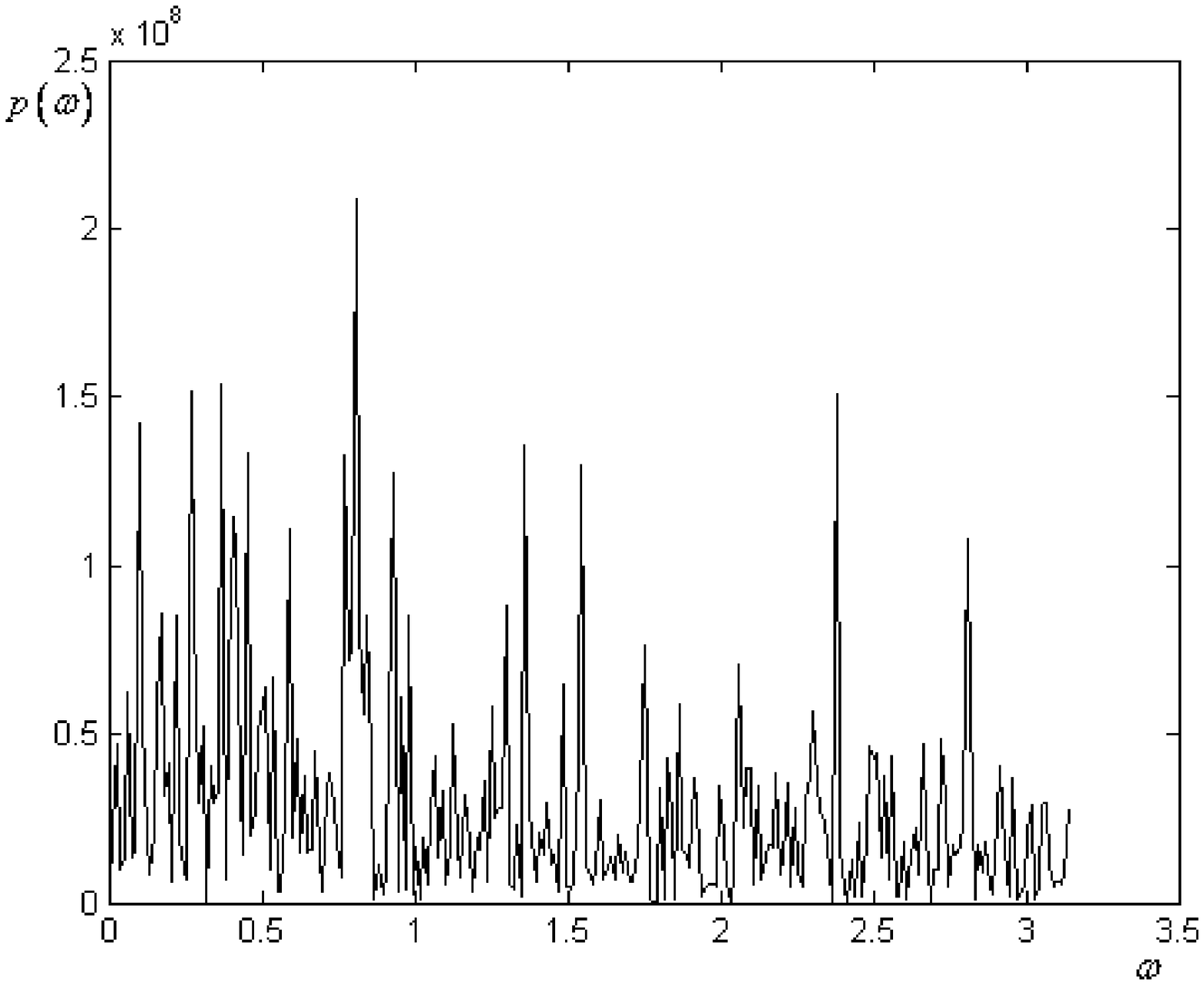}
  \caption{Fourier image of correlation function $G_{p}(\tau)=\langle P(t+\tau)P(t)\rangle$ ,$G_{p}(\omega)=\int d\tau
exp(i\omega \tau)G_{p}(\tau)$. Finite width of correlation
function confirms the existence of chaos. The graph is plotted for
the same values of the parameters as for Fig.1.}\label{fig:2}
\end{figure}

The other parameters of the system have also chaotic spectrum (see Fig.3).
\begin{figure}[t]
 \centering
  \includegraphics[width=14cm]{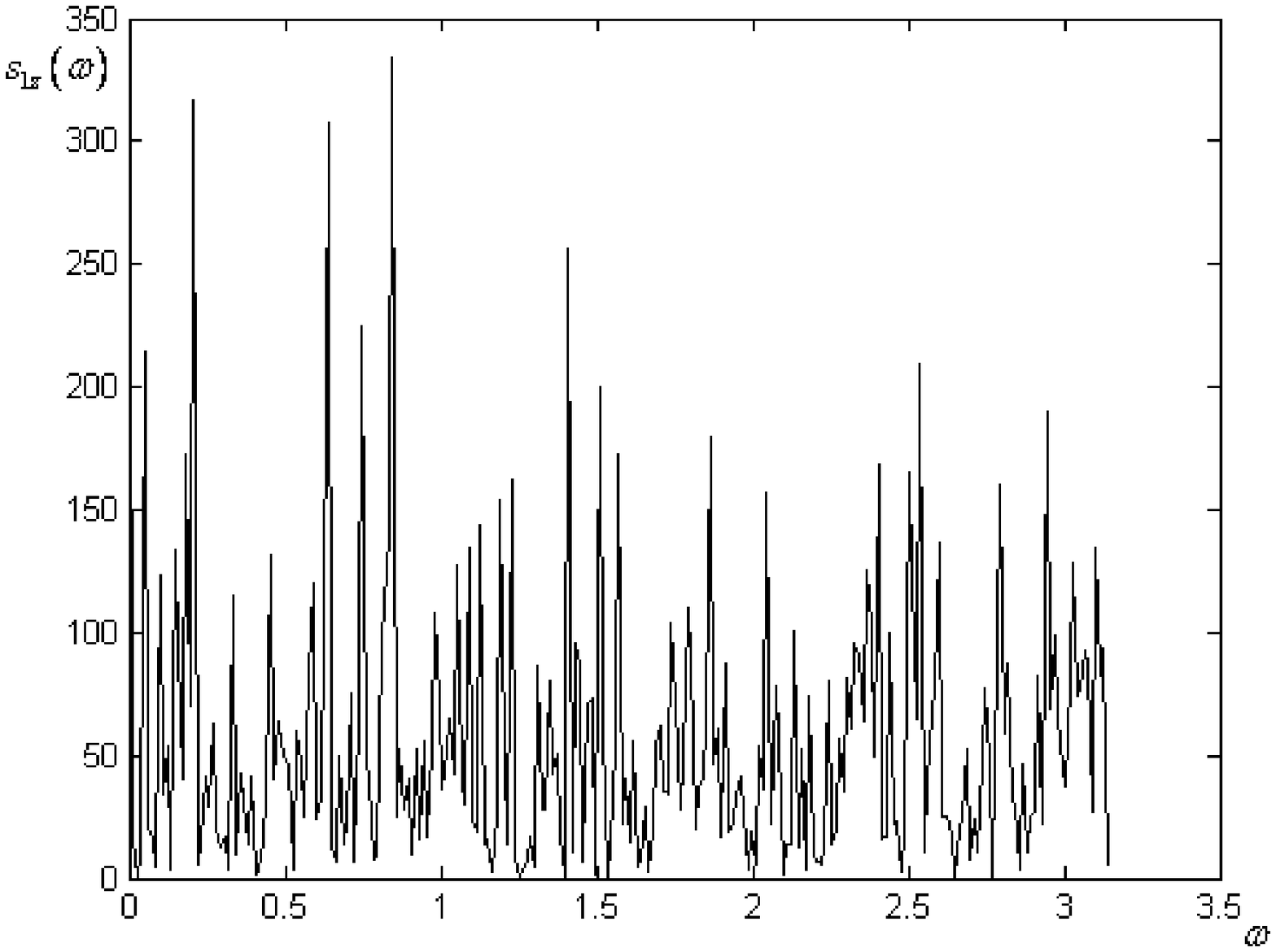}
  \caption{Fourier image of correlation function of variable
$S_{1}^{z}$. The numerical vales of the parameters are analogous
of that of Fig.1.}\label{fig:3}
\end{figure}

In order to be more convinced of dynamics to be chaotic, we have
calculated even fractal dimension of the system.

In order to calculate fractal dimension of the system's phase
space we use the Grassberger-Procaccia algorithm
\cite{Grassberger,Procaccia}. The idea of this algorithm is the
following. Let us suppose, we obtain an ensemble of state vectors
${x_{i},~i=1,2,...N}$ by numerical solving of the set of
equations, corresponding to successive steps of integration of
differential equations. Choosing small parameter $\varepsilon$ we
can use our result for evolution of the following sum: \be
C(\varepsilon)=\lim_{N\rightarrow\infty}\frac{1}{N(N-1)}\sum_{i,j=1}^{N}\theta(\varepsilon-|x_{i}-x_{j}|),\ee
where $\theta$ is a step function \be \theta(x)=\left\{
\begin{array}{l}
0\hskip0.5cm x<0\\
1\hskip0.5cm x\geq0\\\end{array}\right .\ee

According to Grassberger-Procaccia algorithm, if we know
$C(\varepsilon)$, we can estimate strange attractor's fractal
dimension with the help of the following formula
\cite{Grassberger,Procaccia} \be D =
\frac{C(\varepsilon)}{\lg(\varepsilon)}.\ee

The numerical results are represented on Fig.4.
\begin{figure}[t]
 \centering
  \includegraphics[width=14cm]{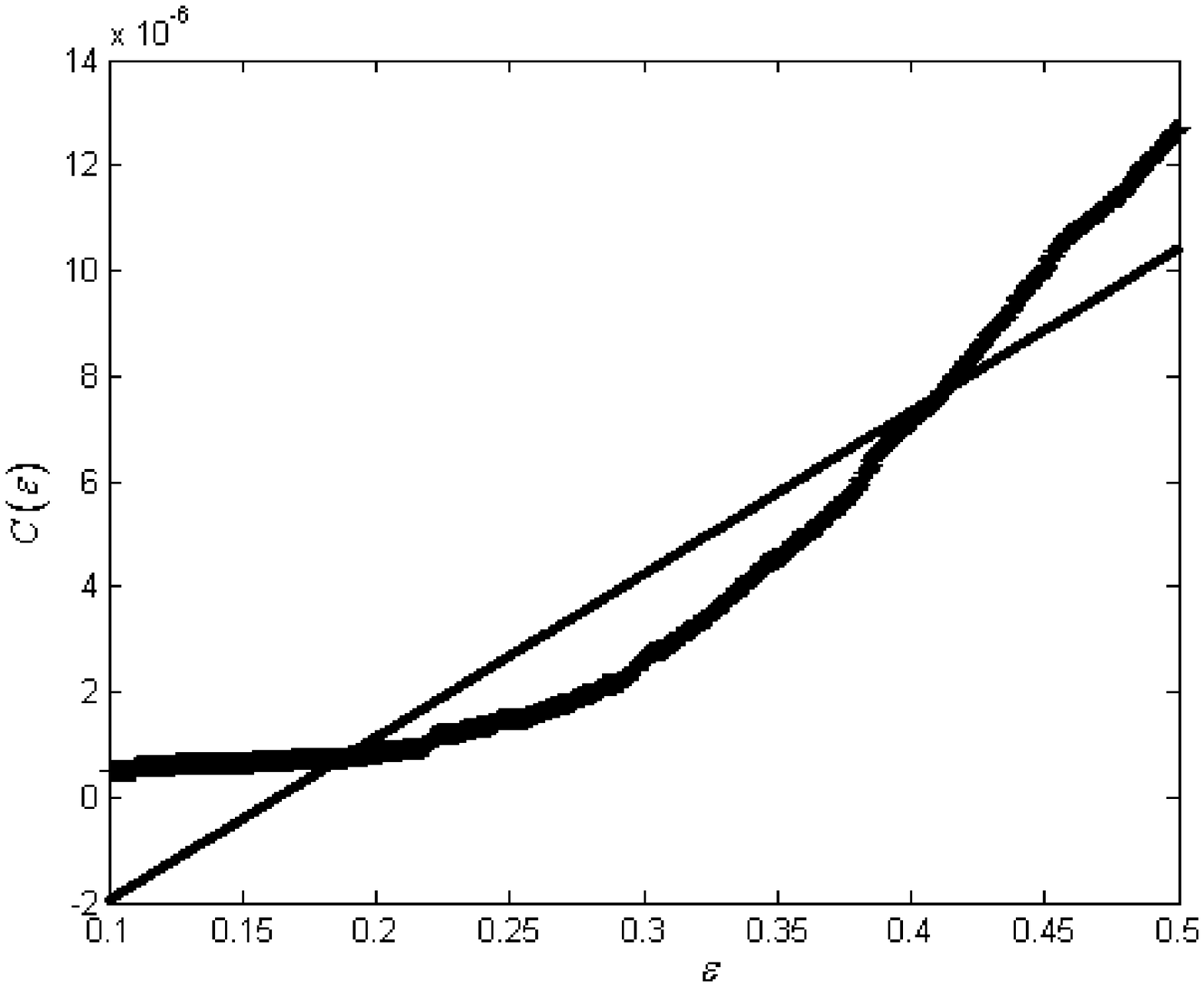}
  \caption{The graph of dependence of $C(\varepsilon)$ on $\varepsilon$
plotted using Grassberger-Proccacia algorithm for the values of
the parameters analogous of that of Fig.1. A solid line
corresponds to least-squares approximation of the results of date
processing. The estimated fractal dimension is equal to
$D=\frac{\ln(C(\varepsilon_{
2}))-ln(C(\varepsilon_{1}))}{\ln\varepsilon_{2}-\ln\varepsilon_{1}}\approx
2.2$,$(\varepsilon_{1}\approx0.12,
\varepsilon_{2}\approx0.41,C(\varepsilon_{1})\approx0.5\cdot10^{-6},C(\varepsilon_{2})\approx7.5\cdot10^{-6})$.
The numerical data obtained verify that the dynamics of the system
is chaotic. We shall make use of this fact in the second part of
this work where quantum-statistical description will be used for
the study of the systems dynamic without use of quasi-classical
methods.}\label{fig:4}
\end{figure}

II. As we have shown in the first part of the work the dynamics of
the system is chaotic for certain values of parameters even for
zero detuning. When considering the state of the system with
quantum-statistical methods we shall neglect kinetic energy of the
system and operator $\hat{x}$ will be regarded as classical
chaotic variable $x(t)$ , presented itself stochastic process.
Condition of using this kind of approximation is the following:
Acting on the system classical force is
$$|\overrightarrow{F}|=\frac{\Delta P}{\Delta t}=|\nabla_{x}\hat{H}_{SB}|\approx g_{0}K_{f}$$
So, classical momentum transferred to the atom is $\Delta
P\approx\Delta tg_{0}K_{f}$. Then influence of the atomic motion
on the energy levels can be neglected if $\frac{(\Delta
P)^{2}}{2m}<|\hat{H}_{SB}|.$ \\This means
$\frac{g_{0}K_{f}^{2}(\Delta t)^{2}}{2m}<1$.

Let us write Schrodinger equation of the system in interaction
representation: \be i\frac{d|\psi(t)>}{dt}=\hat{V}|\psi(t)>,\ee
where \be
\hat{V}=\Omega(\hat{S}_{1}^{+}\hat{S}_{2}^{-}+\hat{S}_{1}^{-}\hat{S}_{2}^{+})+\omega_{f}\hat{b}^{+}\hat{b}-g_{0}\cos(k_{f}\hat{x})((\hat{S}_{1}^{+}+
\hat{S}_{2}^{+})\hat{b}+(\hat{S}_{1}^{-}+\hat{S}_{2}^{-})\hat{b}^{+})\ee
is an interaction operator.

Assume that at zero time $t=0$, the system's wave function
represents itself direct product of wave functions of atom
$|\psi_{atom}>$ and $|\psi_{field}>$ field:
$$ |\psi(t=0)>=|\psi_{atom}>\otimes|\psi_{field}>.$$
Here \be |\psi_{atom}> = C_{00}|00> + C_{01}|01> + C_{10}|10> +
C_{11}|11>,\ee \be |\psi_{field}>=\sum_{n}W_{n}|n>,\ee where
$|n,m>$ is qubit's wave function.

Because of interaction (16) the following transition between
states are possible: \be |0,0,n+1>\leftrightarrow
|0,1,n>,~~|0,0,n+1>\leftrightarrow |1,0,n>,\ee \be
|0,1,n>\leftrightarrow |1,1,n-1>,~~|1,0,n>\leftrightarrow
|1,1,n-1>,~~|1,0,n>\leftrightarrow |0,1,n>.\ee

The transition (19) correspond to the transitions between energy
states with changing number of photons and the transitions (20)
correspond to inter spin transitions. On the basis of equations
(19),(20) we shall search for the solution of equation (15) in the
following form:
$$|\psi(t)>=\sum_{n}C_{0,0,n+1}|0,0,n+1>+\sum_{n}C_{0,1,n}|0,1,n>+$$
\be +\sum_{n}C_{1,0,n}|1,0,n>+\sum_{n}C_{1,1,n-1}|1,1,n-1>.\ee
Taking into account equations (15)-(21), we obtain the following
equations for coefficients of resolution:
$$i\frac{dC_{0,0,n+1}}{dt} = \Omega C_{0,0,n+1}-g_{0}\cos(k_{f}x(t))\sqrt{n+1}(C_{1,0,n}+C_{0,1,n}),$$
$$i\frac{dC_{1,1,n-1}}{dt} = \Omega C_{1,1,n-1}-g_{0}\cos(k_{f}x(t))\sqrt{n}(C_{1,0,n}+C_{0,1,n}),$$
$$i\frac{dC_{0,1,n}}{dt} = \Omega C_{1,0,n}-g_{0}\cos(k_{f}x(t))(C_{1,1,n-1}\sqrt{n}+C_{0,0,n+1}\sqrt{n+1}),$$
\be i\frac{dC_{1,0,n}}{dt} = \Omega
C_{0,1,n}-g_{0}\cos(k_{f}x(t))(C_{1,1,n-1}\sqrt{n}+C_{0,0,n+1}\sqrt{n+1}).\ee
In the set of equations (22) let us pass to the new variables: \be
A(t)=C_{1,0,n}+C_{0,1,n},~~B(t)=\sqrt{n+1}C_{0,0,n+1}+\sqrt{n}C_{1,1,n-1}.\ee
Taking into account (23), the set (22) takes the following form:
$$ i\frac{d A(t)}{dt}=\Omega
A(t)-2g_{0}\cos(k_{f}x(t))B(t),$$ \be i\frac{d B(t)}{dt}=\Omega
B(t)-(2n+1)g_{0}\cos(k_{f}x(t))A(t).\ee If we introduce the new
notations: \be
B^{\prime}(t)=\sqrt{2}B(t),~~A^{\prime}(t)=\sqrt{2n+1}A(t),~~\omega(t)=\sqrt{2}\sqrt{2n+1}g_{0}\cos(k_{f}x(t))\ee
and after that: \be
C(t)=A^{\prime}(t)+B^{\prime}(t),~~D(t)=A^{\prime}(t)-B^{\prime}(t).\ee
Taking into account equations (25) and (26), the set of equations
(24) takes the simpler form:
$$i\frac{dC(t)}{dt}=\Omega C(t)-\omega(t)C(t),$$
\be i\frac{dD(t)}{dt}=\Omega D(t)+\omega(t)D(t).\ee It is readily
seen that the solutions of the set (27) have the following form:
\be C(t)=C_{1}e^{i\Omega t}e^{i\int_{0}^{t}\omega
(t^{\prime})dt^{\prime}},~~D(t)=C_{2}e^{i\Omega
t}e^{-i\int_{0}^{t}\omega (t^{\prime})dt^{\prime}}.\ee Let us
introduce the notations for the functionals: \be
Q[\omega(t)]=e^{i\int_{0}^{t}\omega (t^{\prime})dt^{\prime}},\ee
\be Q^{\ast}[\omega(t)]=Q^{-1}[\omega(t)]=e^{-i\int_{0}^{t}\omega
(t^{\prime})dt^{\prime}}.\ee Taking into account (27)-(30), the
solutions of the set (24) takes the form: \be
A(t)=\frac{A^{\prime}(t)}{\sqrt{2n+1}}=\frac{C_{1}}{2\sqrt{2n+1}}e^{-i\Omega
t}Q[\omega(t)]+\frac{C_{2}}{2\sqrt{2n+1}}e^{-i\Omega
t}Q^{-1}[\omega(t)],\ee \be
B(t)=\frac{B^{\prime}(t)}{\sqrt{2}}=\frac{C_{1}}{2\sqrt{2}}e^{-i\Omega
t}Q[\omega(t)]-\frac{C_{2}}{2\sqrt{2}}e^{-i\Omega
t}Q^{-1}[\omega(t)].\ee Taking into account (31), (32) and (23) we
obtain: \be
C_{1,0,n}+C_{0,1,n}=\frac{C_{1}}{2\sqrt{2n+1}}e^{-i\Omega
t}Q[\omega(t)]+\frac{C_{2}}{2\sqrt{2n+1}}e^{-i\Omega
t}Q^{-1}[\omega(t)],\ee \be
\sqrt{n+1}C_{0,0,n+1}+\sqrt{n}C_{1,1,n-1}=\frac{C_{1}}{2\sqrt{2}}e^{-i\Omega
t}Q[\omega(t)]-\frac{C_{2}}{2\sqrt{2}}e^{-i\Omega
t}Q^{-1}[\omega(t)].\ee The equations (33) and (34) are the
conditions to determine time dependence of the coefficients of the
functions (21). But for determination of four coefficients we need
two more conditions. The third condition for determination of
coefficients $C_{1,0,n}(t)$ and $C_{0,1,n}(t)$ is easily obtained
from equation (22) and has the following form: \be
i\frac{d(C_{0,1,n}-C_{1,0,n})}{dt}=-\Omega(C_{0,1,n}-C_{1,0,n}).\ee
From this we have: \be C_{0,1,n}-C_{1,0,n}=C_{3}e^{i\Omega t}.\ee
In order to obtain the last fourth condition, we introduce the
notation: \be \sqrt{n+1}C_{0,0,n+1}-\sqrt{n}C_{1,1,n-1}=F(t).\ee
Then taking into account (22) we obtain for $F(t)$: \be
i\frac{dF(t)}{dt}=\Omega F(t)-g_{0}\cos(k_{f}x(t))A(t).\ee The
solution (38) has the form:
$$F(t)=\frac{iC_{1}e^{-i\Omega
t}}{2\sqrt{2}(2n+1)}\int\limits_{0}^{t}\omega(t^{\prime})Q[\omega(t^{\prime})]dt^{\prime}+$$
\be +\frac{iC_{2}e^{-i\Omega
t}}{2\sqrt{2}(2n+1)}\int\limits_{0}^{t}\omega(t^{\prime})Q^{-1}[\omega(t^{\prime})]dt^{\prime}+C_{4}e^{-i\Omega
t}.\ee For further simplification of equation (39) consider the
expression: \be
\int\limits_{0}^{t}\omega(t^{\prime})Q[\omega(t^{\prime})]dt^{\prime}=
\int\limits_{0}^{t}\omega(t^{\prime})e^{i\int\limits_{0}^{t^{\prime}}\omega(t^{\prime\prime})dt^{\prime\prime}}dt^{\prime}\ee
and let us introduce the notation: \be \Omega_{0}
(t^{\prime})=\int\limits_{0}^{t^{\prime}} \omega
(t^{\prime\prime})dt^{\prime\prime}.\ee Then it is readily seen
that: \be
\int\limits_{0}^{t}\omega(t^{\prime})Q[\omega(t^{\prime})]dt^{\prime}=\int\limits_{0}^{t}d\Omega_{0}(t^{\prime})e^{i\Omega_{0}(t^{\prime})}=-i(e^{i\Omega_{0}(t)}-1).\ee
By analogy with previous one: \be
\int\limits_{0}^{t}\omega(t^{\prime})Q^{-1}[\omega(t^{\prime})]dt^{\prime}=\int\limits_{0}^{t}d\Omega_{0}(t^{\prime})e^{-i\Omega_{0}(t^{\prime})}=i(e^{-i\Omega_{0}(t)}-1).\ee
Taking into account (42), (43), the expression (39) takes the
form: \be F(t)=\frac{C_{1}e^{-i\Omega
t}}{2\sqrt{2}(2n+1)}(Q[\omega(t)]-1)-\frac{C_{2}e^{-i\Omega
t}}{2\sqrt{2}(2n+1)}(Q^{-1}[\omega(t)]-1)+C_{n}e^{-i\Omega t}.\ee
Taking into consideration (33), (34),(36) and (44) we can yet
write down the set of four algebraic equations for the
coefficients of wave function (21):
$$C_{1,0,n}+C_{0,1,n}=\frac{C_{1}}{2\sqrt{2n+1}}e^{-i\Omega
t}Q[\omega(t)]+\frac{C_{2}}{2\sqrt{2n+1}}e^{-i\Omega
t}Q^{-1}[\omega(t)],$$
$$\sqrt{n+1}C_{0,0,n+1}+\sqrt{n}C_{1,1,n-1}=\frac{C_{1}}{2\sqrt{2}}e^{-i\Omega
t}Q[\omega(t)]-\frac{C_{2}}{2\sqrt{2}}e^{-i\Omega
t}Q^{-1}[\omega(t)],$$
$$C_{0,1,n}-C_{1,0,n}=C_{3}e^{i\Omega t},$$
$$\sqrt{n+1}C_{0,0,n+1}-\sqrt{n}C_{1,1,n-1}=$$ \be =\frac{C_{1}e^{i\Omega
t}}{2\sqrt{2}(2n+1)}(Q[\omega(t)]-1)-\frac{C_{2}e^{-i\Omega
t}}{2\sqrt{2}(2n+1)}(Q^{-1}[\omega (t)]-1)+C_{4}e^{-i\Omega t}.\ee

Here the coefficients of integration are connected with the
initial conditions via the relations:
$$C_{1}=\sqrt{2n+1}(C_{0,1,n}(0)+C_{1,0,n}(0))+\sqrt{2}(\sqrt{n+1}C_{0,0,n+1}(0)+\sqrt{n}C_{1,1,n-1}(0)),$$
$$C_{2}=\sqrt{2n+1}(C_{0,1,n}(0)+C_{1,0,n}(0))-\sqrt{2}(\sqrt{n+1}C_{0,0,n+1}(0)+\sqrt{n}C_{1,1,n-1}(0)),$$
$$C_{3}=C_{0,1,n}-C_{1,0,n},$$
\be C_{4}=C_{0,0,n+1}\sqrt{n+1}-C_{1,1,n-1}\sqrt{n}.\ee  by
solving the set of equations (45), it is possible to determine
time dependence of wave function (21) and by means of this to
determine quantum state of qubit:
$$C_{0,0,n+1}(t)= \frac{C_{1}e^{-i\Omega
t}\sqrt{n+1}}{2\sqrt{2}(2n+1)}Q[\omega(t)]-\frac{C_{2}e^{-i\Omega
t}\sqrt{n+1}}{2\sqrt{2}(2n+1)}Q^{-1}[\omega(t)]+$$
$$+(\frac{C_{4}}{2\sqrt{n+1}}-\frac{C_{1}}{4\sqrt{2}\sqrt{n+1}(2n+1)}+
\frac{C_{2}}{4\sqrt{2}\sqrt{n+1}(2n+1)})e^{-i\Omega t},$$
$$C_{1,1,n-1}(t)= \frac{C_{1}e^{-i\Omega
t}}{2\sqrt{2}}\frac{\sqrt{n}}{(2n+1)}Q[\omega(t)]-\frac{C_{2}e^{-i\Omega
t}}{2\sqrt{2}}\frac{\sqrt{n}}{(2n+1)}Q^{-1}[\omega(t)]+$$
$$+(-\frac{C_{4}}{2\sqrt{n+1}}+\frac{C_{1}}{4\sqrt{2}\sqrt{n+1}(2n+1)}-
\frac{C_{2}}{4\sqrt{2}\sqrt{n+1}(2n+1)})e^{-i\Omega t},$$
$$C_{1,0,n}(t)=\frac{C_{1}}{4\sqrt{2n+1}}e^{-i\Omega
t}Q[\omega(t)]+\frac{C_{2}}{4\sqrt{2n+1}}e^{-i\Omega
t}Q^{-1}[\omega(t)]+\frac{C_{3}}{2}e^{i\Omega t},$$ \be
C_{0,1,n}(t)=\frac{C_{1}}{4\sqrt{2n+1}}e^{-i\Omega
t}Q[\omega(t)]+\frac{C_{2}}{4\sqrt{2n+1}}e^{-i\Omega
t}Q^{-1}[\omega(t)]-\frac{C_{3}}{2}e^{i\Omega t}.\ee

As is seen from (47), time dependence of the coefficients of wave
function (21) describing quantum state of qubit is determined by
the functional: \be
Q[\omega(t)]=e^{i\int_{0}^{t}\omega(t^{\prime})dt^{\prime}},\ee
where \be \omega(t)=\sqrt{2(2n+1)}g_{0}\cos(k_{f}x(t)).\ee As is
seen from (49) time dependence of quantum state depends on $x(t)$.
Thus in order to determine qubit's state, it is necessary to know
the coordinate of the, system as explicit function of time $x(t)$.
But on the other hand as we have showed in the first part of the
work, because of the dynamics to be chaotic $x(t)$ may be
considered as classical chaotic process. In this case for
determination of the system's state it is necessary to average the
functional (48) by all realizations of stochastic variable $x(t)$.
For this, we represent stochastic average of functional (48) in
the form of the following continual integral: \be
<Q[\omega(t)]>=<exp(i\int_{0}^{t}\omega
(t^{\prime})dt^{\prime})>=\lim\limits_{N\rightarrow\infty \atop
\Delta t_{k}\rightarrow0}\int d\omega_{N} \ldots
d\omega_{1}exp(i\sum_{k=i}^{N}\omega_{k}\Delta
t_{k})P_{N}(\omega),\ee where \be P_{N}(\omega)=(2\pi)^{-N} \int
d\lambda_{1}\ldots
d\lambda_{N}exp[-i\sum_{k=i}^{N}\lambda_{k}\omega_{k}]exp[-\frac{1}{2}\sum_{kk^{\prime}}C_{kk^{\prime}}\lambda_{k}\lambda_{k^{\prime}}]\ee
is Fourier image of distribution function, $\Delta
t_{k}=t^{(k)}-t^{(k-1)},~t^{(0)}=0,~t^{(N)}=t.$

It is readily seen that by taking into account (51), the
expression (50) can be rewritten in the following form:
$$\int d\omega_{N}\ldots
d\omega_{1}exp(i\sum_{k=1}^{N}\omega_{k}\Delta
t_{k})P_{N}(\omega)=$$
$$=\int d\lambda_{1}\ldots d\lambda_{N} exp[-\frac{1}{2}\sum_{kk^{\prime}}C_{kk^{\prime}}\lambda_{k}\lambda_{k^{\prime}}]\prod_{k=1}^{N}\frac{1}{2\pi}\int exp[i\omega_{k}(\Delta t_{k}-\lambda_{k})]=$$
\be =\int d\lambda_{1}\ldots d\lambda_{k}\delta
(\lambda_{1}-\Delta t_{1})\delta (\lambda_{2}-\Delta t_{2})\ldots
\delta (\lambda_{N}-\Delta
t_{N})exp[-\frac{1}{2}\sum_{kk^{\prime}}C_{kk^{\prime}}\lambda_{k}\lambda_{k^{\prime}}].\ee
Taking into account (52)  for statistically averaged functional we
obtain:
$$<Q[\omega(t)]>=\lim_{N\rightarrow\infty}exp[-\frac{1}{2}\sum_{kk^{\prime}}C(t^{(k)},t^{(k^{\prime})})\Delta t^{k}\Delta t^{k^{\prime}}]=$$
\be
=exp(-\frac{1}{2}\int\limits_{0}^{t}dt^{\prime}\int\limits_{0}^{t}d
t^{\prime\prime}C(t^{\prime},t^{\prime\prime})).\ee For random
processes
$C(t^{\prime},t^{\prime\prime})=C(t^{\prime}-t^{\prime\prime})$.
Then introducing the new variables:
$t^{\prime}-t^{\prime\prime}=\tau$,
$t^{\prime}+t^{\prime\prime}=\xi$, and assuming that correlation
function has Gaussian form \\
$C(\tau)=<\omega(t+\tau)\omega(\tau)>=e^{-\alpha_{0}\tau^{2}}$,
finally from (53) we obtain: \be
<Q[\omega(t)]>=exp[-\frac{t}{2}\sqrt{\frac{\pi}{\alpha_{0}}}Erf(t\sqrt{\alpha_{0}})],\ee
where $Erf( \ldots )$ is error function \cite{Handbook}.

Assume that at zero time the system was in the state: \be
|\psi(0)>=|\psi_{atom}> \otimes |\psi_{field}>,\ee where \be
|\psi_{field}>=\sum _{n} W_{n}|n>.\ee Comparing (55), (56) with:
$$|\psi(0)>=\sum_{n}C_{0,0,n+1}(0)|0,0,n+1>+\sum_{n}C_{0,1,n}(0)|0,0,n>+$$
\be +
\sum_{n}C_{1,0,n}(0)|1,0,n>+\sum_{n}C_{1,1,n-1}(0)|1,1,n-1>\ee it
is possible to obtain the following relations for the initial
conditions:
$$C_{00}W_{n+1}=C_{00n+1}(0),~~C_{01}W_{n}=C_{01n}(0),$$
\be C_{10}W_{n}=C_{10n}(0),~~C_{00}W_{n+1}=C_{00n+1}(0).\ee Let us
determine the values measured on experiment that are connected
with population difference of levels:
$$I_{11,01}=W(t,|11>)-W(t,|01>),$$
$$I_{11,10}=W(t,|11>)-W(t,|10>),$$
$$I_{10,00}=W(t,|10>)-W(t,|00>),$$
$$I_{01,00}=W(t,|01>)-W(t,|00>),$$
\be
I_{11,00}=W(t,|11>)-W(t,|00>)=\frac{1}{2}(I_{11,01}+I_{11,10})+\frac{1}{2}(I_{01,00}+I_{01,00}),\ee
where
$$W(t,|11>)=\sum_{n=0}^{\infty}|C_{1,1,n-1}(t)|^{2},$$
$$W(t,|01>)=\sum_{n=0}^{\infty}|C_{0,1,n}(t)|^{2},$$
$$W(t,|10>)=\sum_{n=0}^{\infty}|C_{1,0,n}(t)|^{2},$$
\be W(t,|00>)=\sum_{n=0}^{\infty}|C_{0,0,n+1}(t)|^{2}.\ee For
illustration let us calculate for example $W(t,|1,0>)$. Taking
into account (47) and (54) we obtain:
$$<W(t,|10>)>=\frac{1}{8}\sum_{n=0}^{\infty}(C_{10}W_{n}+C_{01}W_{n})^{2}+\frac{1}{4}\sum_{n=0}^{\infty}\frac{1}{2n+1}(\sqrt{n+1}C_{0,0}W_{n+1}+$$
\be
+\sqrt{n}C_{1,1}W_{n-1})^{2}+\frac{1}{4}\sum_{n=0}^{\infty}(C_{1,0}W_{n}-C_{0,1}W_{n})^{2}+
<W(t,|10>)>_{int},\ee where  $<W(t,|1,0>)>_{int}$ denote
interference terms whose explicit forms are not brought here for
the sake of brevity. The point is that interference terms contain
terms of the following form: \be
<Q^{-2}[\omega(t)]>,~~<Q^{2}[\omega(t)]>,~~<e^{2i\Omega
t}Q^{-1}[\omega(t)]>.\ee These quantities, as well as (54), fall
down quickly after the lapse of time. For example: \be
<e^{2i\Omega t}Q^{-1}[\omega(t)]>=e^{2i\Omega
t}exp(-\frac{t}{2}\sqrt{\frac{\pi}{\alpha_{0}}}Erf(t\sqrt{\alpha_{0}})).
\ee As is seen from (63), for time interval that is more then the
time of correlation function of the random quantity $\omega(t)$
(49), $t>\sqrt{\frac{\pi}{\alpha_{0}}}$ \be
C(\tau)=<\omega(t+\tau)\omega(\tau)>=e^{-\alpha_{0}\tau^{2}}\ee
interferentional terms can be neglected in (61). Situation is
analogous for other quantities as well from (59),(60). Thus we
were able to prove that because of dynamics to be chaotic zeroing
of interferentional terms occurs. This fact of zeroing of
inerferentional terms has deep physical sense. This means that the
system execute transition from pure quantum-mechanical state to
mixed one \cite{Landau}. Such a transition is irreversible, as
information about the phase of the system is lost. Transition from
pure quantum state to mixed one is one of the manifestations of
quantum chaos
\cite{Ugulava,Chotorlishvili,Nickoladze,Gvarjaladze,Skrinnikov}.
Formulae analogous to (61) can be obtained for other quantities
(60) as well:
$$<W(t,|0,0>)>=\sum_{n=0}^{\infty}\frac{n+1}{4(2n+1)}(C_{10}W_{n}+C_{01}W_{n})^{2}+\frac{1}{2}\sum_{n=0}^{\infty}\frac{n+1}{(2n+1)^{2}}(\sqrt{n+1}C_{0,0}W_{n+1}+$$
$$+\sqrt{n}C_{1,1}W_{n-1})^{2}+\sum_{n=0}^{\infty}\frac{n}{(2n+1)^{2}}(\sqrt{n}C_{0,0}W_{n+1}-\sqrt{n+1}C_{1,1}W_{n-1})^{2}$$
$$<W(t,|11>)>=\sum_{n=0}^{\infty}\frac{n}{4(2n+1)}(C_{10}W_{n}+C_{01}W_{n})^{2}+\frac{1}{2}\sum_{n=0}^{\infty}\frac{n}{(2n+1)^{2}}(\sqrt{n+1}C_{0,0}W_{n+1}+$$
$$+\sqrt{n}C_{1,1}W_{n-1})^{2}+\sum_{n=0}^{\infty}\frac{n+1}{(2n+1)^{2}}(\sqrt{n}C_{0,0}W_{n+1}-\sqrt{n+1}C_{1,1}W_{n-1})^{2}$$
$$<W(t,|10>)>=<W(t,|0,1>)>=\frac{1}{8}\sum_{n=0}^{\infty}(C_{10}W_{n}+C_{01}W_{n})^{2}+$$
\be
+\frac{1}{4}\sum_{n=0}^{\infty}\frac{1}{2n+1}(\sqrt{n+1}C_{0,0}W_{n+1}+\sqrt{n}C_{1,1}W_{n-1})^{2}+\frac{1}{4}\sum_{n=0}^{\infty}(C_{10}W_{n}-C_{01}W_{n})^{2}\ee
where $C_{00},~C_{01},~C_{10},~C_{11}$ quantities are populations
of corresponding levels, $W_{n}$ describes state of the field. It
is usually assumed that $W_{n}$ satisfy Gaussian distribution
\cite{Schleich}: \be W_{n}=\frac{1}{\sqrt{2\pi\Delta
n^{2}}}exp[-\frac{(n-\bar{n}) ^{2}}{\Delta n^{2}}].\ee As we noted
above transition from pure state to mixed one is irreversible. In
order this fact to be confirmed, let us calculate change of the
system's entropy.

Let us assume, that the system at zero time was in state
$C_{00,n+1}$ . In this case the system's entropy according to
\cite{Ropke}, \cite{Fujita}  is: \be
S(t=0)=-\sum_{i=1}^{4}\rho_{i}\ln\rho_{i}=0,\ee as only one of the
elements of density matrix $\rho$ is nonzero: \be
\rho_{1}(t=0)=|C_{00,n+1}(0)|^{2}=1,~~\rho_{2}(t=0)=\rho_{3}(t=0)=\rho_{4}(t=0)=0.\ee
After the lapse of time that is more than the time of transition
between the levels $t_{0}\sim1/g_{0}, t>t_{0}$ the system has time
to execute transition between levels. That is why probability to
find system in other states will be nonzero:
 \be C_{11}\neq 0,~~C_{01}\neq
0,~~C_{10}\neq 0~~~~~t>t_{0}.\ee Despite of this fact to talk
about probability of population of different states is early yet.
The point is that in time interval: \be
t_{0}<t<\sqrt{\frac{\pi}{\alpha_{0}}}\ee interferentional terms in
equations (61),(65) are nonzero. Therefore the state of the system
will be pure one. But unlike of the initial state (68),which is
simple state, the state of the system in time interval (70) is
superposition one.

Superposition state is pure quantum mechanical state and only
after zeroing of interferentional terms in (61) and (65)
superposition state passes to mixed one. Such a transition occurs
in times: \be t>\sqrt{\frac{\pi}{\alpha_{0}}}\ee But in time
interval (70) while the system is in pure superposition state,
from the symmetry point of view, it is clear that the coefficient
values(69) have to satisfy the following relation:
$$C_{00}\left(t_{0}<t<\sqrt{\frac{\pi}{\alpha}}\right)\sim
C_{11}\left(t_{0}<t<\sqrt{\frac{\pi}{\alpha}}\right)\sim$$ \be
\sim C_{01}\left(t_{0}<t<\sqrt{\frac{\pi}{\alpha}}\right)\sim
C_{10}\left(t_{0}<t<\sqrt{\frac{\pi}{\alpha}}\right)\sim C.\ee
Taking into account normalization condition: \be
\sum_{i,j=0}^{1}<W(t,|ij>)>=1\ee and (72),  from (65)we obtain:
\be
C^{2}(\sum_{n=0}^{\infty}(W_{n}^{2}+W_{n+1}^{2}+W_{n-1}^{2}))=1.\ee
Then taking into account the relation: \be
<W(t,|0,1>)>=<W(t,|1,0>)>\ee it is easy to obtain the condition
from (65):
$$<W(t,|11>)>+<W(t,|00>)>=<W(t,|01>)>+<W(t,|10>)>+$$
\be
(+\sum_{n=0}^{\infty}\frac{n}{(2n+1)}(\sqrt{n}C_{00}W_{n+1}-\sqrt{n+1}C_{11}W_{n-1})^{2}.\ee
The condition (76) means in its turn that at times (71) mixed
states are formed in the system in which the levels:  \be
\rho_{1}=<W(t>\sqrt{\pi/\alpha}|11>)>=a,~~\rho_{2}=<W(t>\sqrt{\pi/\alpha}|00>)>=b\ee
are populated with more probability than the levels: \be
\rho_{3}=<W(t>\sqrt{\pi/\alpha}|01>)>=\rho_{4}=<W(t>\sqrt{\pi/\alpha}|10>)>=c,\ee
where quantities $a,b,c$ satisfy normalization condition: \be
Tr(\hat{\rho})=a+b+2c=1,~~a+b>2c\ee Taking into account (67),
(77), (78) and (79) it is easy to see that during evolution of the
system from pure quantum-mechanical state (68) to mixed one (77),
(78) increase of entropy occurs. $$\Delta
S=S(t>\sqrt{\pi/\alpha})-S(t=0)=-(a\ln a+b\ln b+2c\ln c )>0,$$ \be
0<a<1,~~0<b<1,~~0<c<1\ee

\section{Conclusion}

Let sum up and analyze the results obtained in conclusion.

The aim of this work was to study two 1/2 spin qubit system being
subject to resonator field. Interest to such a systems is caused
by the fact that they are the most perspective to be used in
quantum computer. The question that came up is the following: by
how much will be state of the system controllable and dynamics
reversible? We have considered the most general case, when
interaction of the system with field depends on coordinate of the
system inside resonator.

Contrary to generally accepted opinion, it has turned out that the
absence of detuning between resonator field and frequency of the
system does not guarantee reversibility of the system's state.
During evolution in time the system executes irreversible
transition from pure quantum-mechanical state to mixed one. At the
same time, the time needed for formation of mixed state
$t>\sqrt{\pi/\alpha_{0}}$ is determined completely by the system's
parameters $\alpha=\frac{K_{f}^{2}}{mg_{0}}$.

One more peculiarity of the problem studied is the following. It
is well known
\cite{Buchleitner,Saif,Farhan,Perel'man,Leichtle,Averbukh} that
for integrable quantum systems complete and fractional quantum
revivals are typical
\cite{Buchleitner,Saif,Farhan,Perel'man,Leichtle,Averbukh}. In our
case because of dynamics to be chaotic phase incursion occurs.
This results in zeroing of interferentional terms and irreversible
losing of information about the system's state. This guaranties
the absence of quantum revivals for our system. the noted fact may
be especially interesting for experimental investigation of the
system under consideration.

\newpage


\begin{thebibliography}{40}
%
\bibitem{Aoki} T. Aoki et al. Nature \textbf{443}, 671 (2006).
%
\bibitem{Mabuchi} H. Mabuchi and A. Doherty, Science \textbf{298}, 1372 (2002).
%
\bibitem{Hood} C.J. Hood et al., Science \textbf{287}, 1447 (2000).
%
\bibitem{Raimond} J.Raimond, M.Brune, and S.Haroche, Rev.Mod.Phys. \textbf{73}, 565 (2001).
%
\bibitem{Tuchette} Q.A. Tuchette et al.,
Phys.Rev.Lett. \textbf{75}, 4710 (1995).
%
\bibitem{Wineland} D.J. Wineland et al.,
J.Res.Nat.Inst.Stand.Technol. \textbf{103}, 259 (1998).
%
\bibitem{Schleich} P. Schleich, Quantum Optics in Phase Space, Wiley. VCH,
Berlin (2001).
%
\bibitem{Fujisaki} H. Fujisaki, T. Miyadera, and A. Tanaka, Phys.Rev.E
\textbf{67}, 066201 (2003).
%
\bibitem{Scott} A.J. Scott and C.M. Caves, J.Phys. A \textbf{36}, 9553 (2003).
%
\bibitem{Novaes} M. Novaes and Marcus A.M. de Aguiar, Phys.Rev E
\textbf{70},045201(R)(2004).
%
\bibitem{Xie} Q. Xie and W. Hai, Eur.Phys.J. \textbf{33D}, 265 (2005).
%
\bibitem{Angelo} R.M. Angelo, K. Furuya, M.C. Nemes, and G.Q. Rellegrino, Phys.Rev.A
\textbf{64}, 043801 (2001).
%
\bibitem{Ye} J. Ye, D.W. Yemooy, and H.J.
Kimble, Phys.Rev.Lett. \textbf{83}, 4987 (1999).
%
\bibitem{van Enk} S.J. van Enk, J. McKeever, H.J. Kimble, and J. Ye, Phys.Rev.A.
\textbf{64}, 013407(2001).
%
\bibitem{Punstermann} P.M. Punstermann, T. Fischer, P. Maunz,
P.W.H. Pinkse, and G. Rempe, Phys.Rev.Lett. \textbf{82}, 379
(1999).
%
\bibitem{Loss} D. Loss
and D.P. DiVincenzo, Phys.Rev. A \textbf{57}, 120 (1998.)
%
\bibitem{Kane} B.E. Kane, Nature \textbf{393}, 133 (1998).
%
\bibitem{Skinner} A.J. Skinner, M.E. Davenport, and B.E. Kane
Phys.Rev.Lett. \textbf{90}, 087901 (2003).
%
\bibitem{Ladd} T.D. Ladd, J.R. Goldman, F. Yamaguchi, Y. Yamamoto, E. Abe, and
K.M.Itoh, Phys.Rev.Lett. \textbf{89}, 017901 (2002).
%
\bibitem{de Sousa} R. de Sousa, J.D. Delgado, and S. Das Sarma, Phys.Rev. A
\textbf{70}, 052304(2004).
%
\bibitem{Xiao-Zhong Yuan} Xiao-Zhong Yuan, Hsi-Sheng Goan, and Ka-Di Zhu, Phys.Rev. B \textbf{75}, 045331
(2007).
%
\bibitem{Prants} S. Prants, N. Edelman, and G. Zaslavsky, Phys.Rev. E
\textbf{66}, 046222 (2002).
%
\bibitem{Landau} L.D.Landau and E.M. Lifschitz, Quantum
Mechanics, Non-relativistic Theory, Pergamon, Oxford (1977).
%
\bibitem{Grassberger} P. Grassberger, Phys.Lett. A \textbf{97}, 227 (1983).
%
\bibitem{Procaccia} P. Grassberger and I. Procaccia, Physica
D,\textbf{9}, 189 (1983).
%
\bibitem{Handbook} Handbook of Mathematical Functions with Formulas, Graphs, and Mathematical Tables National Bureau of Standards,
Applied Mathematical Series, 55, U.S. Government Printing,
(Washington D.C., 1964).
%
\bibitem{Lifschitz} L.D. Landau and E.M. Lifschitz,
Statistical Mechanics, v.5, (in Russian) (Nauka Moscow 1976).
%
\bibitem{Ugulava} A. Ugulava, L. Chotorlishvili, and K. Nickoladze, Phys.Rev. E
\textbf{68}, 026216 (2003).
%
\bibitem{Chotorlishvili} A. Ugulava, L. Chotorlishvili, and K. Nickoladze, Phys.Rev. E
\textbf{70}, 026219 (2004).
%
\bibitem{Nickoladze} A. Ugulava, L. Chotorlishvili, and K. Nickoladze, Phys.Rev. E
\textbf{71}, 056211 (2005).
%
\bibitem{Gvarjaladze} A. Ugulava, L.Chotorlishvili, T. Gvarjaladze, and S.Chkhaidze,
Mod.Phys. Lett. B, \textbf{21}, 415 (2007).
%
\bibitem{Skrinnikov} L. Chotorlishvili, A. Ugulava, T. Kereselidze, V. Skrinnikov,
Mod.Phys.Lett.B \textbf{21}, 79 (2007).
%
\bibitem{Ropke} G. Ropke, Statistische Mechanik fur das Nichtgleichgewicht VEB Deutscher Verlag der
Wissenschaften, Berlin (1987).
%
\bibitem{Fujita} S. Fujita, Introduction to Non-Equilibrium Quantum Statistical Mechanics W.B.Saunders
Company, Philadelphia-London (1966).
%
\bibitem{Buchleitner}A.Buchleitner, D.Delande, J.Zakrzewski,
Phys.Rep. 368(5), 409,(2002).
%
\bibitem{Saif} F. Saif, Physics Reports \textbf{419}, 207 (2005).
%
\bibitem{Farhan} Farhan Saif, Physics Reports \textbf{425}, 369 (2006).
%
\bibitem{Perel'man} I. Sh. Averbukh  and N.F. Perel'man, Phys. Lett. A \textbf{139}, 449.
(1989).
%
\bibitem{Leichtle} C. Leichtle, I. Sh. Averbukh, and W. P. Schleich, Phys. Rev. Lett. \textbf{77}, 3999
(1996).
%
\bibitem{Averbukh} C. Leichtle , I. Sh. Averbukh, and W. P. Schleich, Phys. Rev. A \textbf{54}, 5299
(1996).

%
\end{thebibliography}
\end{document}